\documentclass[conference]{IEEEtran}
\IEEEoverridecommandlockouts

\usepackage{cite}
\usepackage{amsmath,amssymb,amsfonts}
\usepackage{algorithmic}
\usepackage{graphicx}
\usepackage{textcomp}
\usepackage{xcolor}
\usepackage{booktabs}
\usepackage{url}
\usepackage{hyperref}
\usepackage{tikz}
\usetikzlibrary{positioning, arrows.meta, decorations.pathreplacing, calc}

\def\BibTeX{{\rm B\kern-.05em{\sc i\kern-.025em b}\kern-.08em
    T\kern-.1667em\lower.7ex\hbox{E}\kern-.125emX}}

\newcommand{\method}{\textsc{ChainLSTM}\xspace}
\usepackage{xspace}

\begin{document}

\title{Chain-Aware Encoding for Microservice Trace Anomaly Detection}

\author{
\IEEEauthorblockN{Yiliu Xu\IEEEauthorrefmark{1},
Ziwei Hong\IEEEauthorrefmark{2},
Zhongheng Yang\IEEEauthorrefmark{3},
Xinjin Li\IEEEauthorrefmark{4}, and
Yu Ma\IEEEauthorrefmark{1}}
\IEEEauthorblockA{\IEEEauthorrefmark{1}Carnegie Mellon University \quad
\IEEEauthorrefmark{2}Lehigh University \quad
\IEEEauthorrefmark{3}Northeastern University \quad
\IEEEauthorrefmark{4}Columbia University}
}

\maketitle

\begin{abstract}
Microservice traces can be structurally anomalous even when every span returns normally---a payment flow that silently skips a risk check looks fine to any per-span monitor. Sequence models like DeepLog address this by predicting the next event, but they treat each API endpoint as a context-free token: the same endpoint reached through different invocation chains is mapped to the same vocabulary entry, even when its normal behavior differs across contexts. We propose encoding each event as an (endpoint, root-to-span invocation chain) pair instead. This simple change has two consequences: unseen chains are flagged without model inference, and next-event predictions become context-conditional, turning subtle path anomalies into clear outliers. We instantiate this idea in \method, a lightweight dual-task LSTM supporting per-event online detection. On the TrainTicket benchmark, \method achieves 94.3\% F1 (+5.3pp over DeepLog) with comparable latency recall and 99.1\% path recall. Case analysis shows that chain-aware encoding shifts median prediction probability on path anomalies from 0.91 to 0.002, suggesting a wider separation margin for threshold-based detection.
\end{abstract}


\section{Introduction}
\label{sec:intro}

In microservice systems~\cite{dapper,opentelemetry}, an anomalous trace can have every individual span return 200~OK with normal latency---yet the trace pattern is structurally wrong. A payment flow that bypasses a risk check due to a reordering bug looks normal to any per-span monitor. Liu et al.~\cite{traceanomaly} report that rule-based methods at WeBank missed 20\% of anomalies for exactly this reason. These \emph{implicit} anomalies---wrong call ordering, missing calls, context-dependent latency shifts---are invisible to status-code checks and fixed thresholds~\cite{aiops_survey}, yet indicate real faults.

Learning-based approaches address this gap but face trade-offs. DeepLog~\cite{deeplog} predicts the next event in a flat endpoint sequence with an LSTM, enabling real-time detection but ignoring the \emph{invocation chain}---the path through which a service is reached. The same endpoint can have different normal behavior depending on its calling context, and collapsing these contexts creates blind spots. TraceAnomaly~\cite{traceanomaly} captures invocation structure via a whole-trace VAE, but requires the complete trace before detection and is computationally heavyweight. Few existing approaches explicitly combine invocation-chain context with real-time incremental detection.

We propose \method, a lightweight LSTM-based approach that bridges 
this gap. Our contributions are: (1)~an \emph{invocation chain-aware 
encoding} that treats each (endpoint, chain) pair as a distinct token, 
enabling detection of structural anomalies not directly exposed by endpoint-only encodings and producing sharper 
context-conditional predictions; (2)~a lightweight online instantiation of this encoding
in a dual-task LSTM with simple path, latency, and length checks,
supporting per-event detection during trace execution; (3)~an evaluation on TrainTicket~\cite{trainticket} 
showing $94.3\%$ F1, outperforming DeepLog by 5.3 points with only
${\sim}$65K parameters. We focus on trace-level implicit anomalies;
our approach complements log-level and metric-level monitoring. In maintenance
settings, such early structural alerts can help detect deployment regressions,
validate service-call evolution, and narrow debugging to the first
context-violating span.


\section{Related Work}
\label{sec:related}

\textbf{Log- and trace-based approaches.} DeepLog~\cite{deeplog} predicts the next log key from a sliding window via LSTM; extensions~\cite{loganomaly,logrobust} add semantic embeddings. Applied to traces, these methods treat each endpoint as context-independent, conflating events that differ in invocation history. TraceAnomaly~\cite{traceanomaly} captures trace structure via a deep Bayesian network but requires complete traces; graph-based methods~\cite{tracegra} apply GNNs at higher cost; TICAD~\cite{ticad} uses LSTMs at the microservice-pair level but does not tokenize the full root-to-span invocation chain or detect out-of-dictionary chains.

\textbf{Positioning.} We identify a representational limitation in endpoint-only sequence models: when the same endpoint is reached through different invocation contexts, an endpoint-only vocabulary cannot distinguish them, producing multi-modal predictions that mask context-dependent anomalies. \method addresses this not by adding path features to a fixed model, but by changing the vocabulary itself: each (endpoint, chain) pair becomes a distinct token, which both sharpens context-conditional predictions and enables out-of-dictionary structural detection unavailable to endpoint-only models.


\section{Method}
\label{sec:method}

\subsection{Problem Formulation and Detection Scope}
\label{subsec:problem}

A distributed trace $T$ records the execution of a single user request across multiple microservices. We represent $T$ as a time-ordered sequence of API call events:
\begin{equation}
T = \langle e_1, e_2, \ldots, e_n \rangle
\end{equation}
where each event $e_i$ corresponds to a span and is characterized by a tuple:
\begin{equation}
e_i = (o_i, c_i, l_i, \Delta_i, p_i)
\end{equation}
Here, $o_i$ is the endpoint, $c_i$ is the invocation chain leading to this event, $l_i$ is the response latency, $\Delta_i$ is the elapsed time since the previous event, and $p_i$ is the span index within the trace, globally z-score normalized using training-set statistics.

\textbf{Definition (Invocation Chain).} Given a span $s$ in trace $T$, the invocation chain of $s$ is the ordered sequence of service operations from the trace entry point to $s$, including $s$ itself. Formally, if $s$ is reached through parent spans $s_0 \rightarrow \cdots \rightarrow s_{k-1} \rightarrow s$, then $c = [o(s_0), \ldots, o(s_{k-1}), o(s)]$ where $o(\cdot)$ extracts the endpoint of a span.

\textbf{Detection Task.} Given the history $\langle e_1, \ldots, e_{i-1} \rangle$, the goal is to determine at each step $i$ whether $e_i$ is anomalous. A trace is flagged as anomalous if any of its events is detected as such, or if the trace as a whole exhibits structural anomaly.

\textbf{Scope.} Our method targets \emph{implicit} anomalies: path anomalies (call order violations, missing calls, unexpected chains) and latency anomalies (deviations conditioned on the invocation chain). We do not target \emph{explicit} anomalies (HTTP 5xx, hard timeouts) which are handled by infrastructure monitoring.

\subsection{Invocation Chain-Aware Feature Encoding}
\label{subsec:encoding}

We address the representational limitation identified in Section~\ref{sec:related} by treating each event as the pair $(o, c)$ of (endpoint, invocation chain), rather than by endpoint alone. This shift in vocabulary granularity has two distinct consequences for anomaly detection, both exploited in our detection procedure (Section~\ref{subsec:detection}).

\textbf{Consequence 1: Structural anomalies become out-of-dictionary tokens.} Consider an endpoint $E$ that is normally invoked through a fixed chain $A \rightarrow B \rightarrow E$. When a path anomaly removes the intermediate span $B$, $E$ is reached directly, with an invocation chain that has never appeared during training. An endpoint-only model such as DeepLog~\cite{deeplog} sees only $E$, a familiar normal endpoint, and does not raise an alarm. By contrast, a model that includes the chain in its vocabulary recognizes the new \emph{(endpoint, chain)} pair as out-of-dictionary and immediately flags the anomaly. Such structural anomalies are not directly detectable by endpoint-only representations without additional context features.

\textbf{Consequence 2: Next-event distributions become sharply context-conditional.} The same vocabulary shift also concentrates the sequence model's predictions. Two events sharing an endpoint but reached through different chains often have different normal continuations. An endpoint-only vocabulary collapses these contexts, producing a multi-modal next-event distribution in which a context-violating call still appears plausible. A chain-aware vocabulary separates the contexts, yielding a more concentrated distribution and turning subtle context-violating events into clear distributional outliers. We quantify this effect in Section~\ref{subsec:case}.

\textbf{Invocation chain dictionary.} From the training set, we extract all unique pairs $(o, c)$ and assign each a unique integer identifier. Let $\mathcal{D}$ denote this dictionary; we reserve index $0$ for an \texttt{<UNK>} token, assigned to any $(o, c)$ pair not seen during training. The \texttt{<UNK>} token serves as the structural detection signal of Consequence~1.

\textbf{Latency normalization.} For each $(o, c) \in \mathcal{D}$, we compute the per-pair mean $\mu_{(o,c)}$ and standard deviation $\sigma_{(o,c)}$ on the training set, and apply z-score normalization:
\begin{equation}
\tilde{l}_i = \frac{l_i - \mu_{(o_i, c_i)}}{\max(\sigma_{(o_i, c_i)},\, \sigma_{\min})}
\end{equation}
where $\sigma_{\min}$ is set to the 5th percentile of per-chain standard deviations in the training set, preventing inflation when a pair has very few training samples. The elapsed time $\Delta_i$ is normalized globally. During training, we clip $\tilde{l}_i$ to $[-5, 5]$ to limit gradient impact from extreme outliers; at detection time, no clipping is applied so that anomalous events retain their full deviation signal.

\textbf{Event vector.} Each event $e_i$ is encoded as:
\begin{equation}
\mathbf{x}_i = [\mathbf{E}_{\text{chain}}(k_i) \,\|\, \tilde{l}_i \,\|\, \tilde{\Delta}_i \,\|\, p_i]
\end{equation}
where $\mathbf{E}_{\text{chain}} \in \mathbb{R}^{|\mathcal{D}| \times d_e}$ is a learnable embedding, $k_i = \mathcal{D}(o_i, c_i)$ is the chain identifier, and $p_i$ is the span index within the trace, globally z-score normalized using training-set statistics.

\subsection{Dual-Task LSTM Model}
\label{subsec:model}

\method uses a 2-layer LSTM~\cite{lstm} to process a sliding window of $h$ encoded events. The hidden state $\mathbf{h}_{i-1}$ at the last time step feeds two prediction heads. The \emph{path head} produces a softmax distribution $\boldsymbol{\pi}_i$ over $\mathcal{D}$ predicting the next event's chain identifier. The \emph{latency head} produces a scalar $\hat{l}_i$ predicting normalized latency.

We jointly train both heads with a multi-task loss:
\begin{equation}
\mathcal{L} = \mathcal{L}_{\text{path}} + \alpha \cdot \mathcal{L}_{\text{lat}}
\end{equation}
where $\mathcal{L}_{\text{path}}$ is cross-entropy and $\mathcal{L}_{\text{lat}}$ is MSE.

\textbf{Why LSTM and Not Heavier Alternatives.} The vocabulary size $|\mathcal{D}|$ in microservice systems is on the order of $10^2$--$10^3$. An LSTM is sufficient and lightweight enough for real-time per-event detection; Transformers~\cite{transformer} offer greater capacity but at higher latency and memory cost for this scale. Whole-trace VAE-based approaches~\cite{traceanomaly} cannot detect anomalies until trace completion; \method's incremental design enables flagging an anomalous span at step $k$ of an $n$-step trace ($k < n$).

\subsection{Online Detection Procedure}
\label{subsec:detection}

\method combines four complementary detection mechanisms; the first three operate per-event, the fourth at trace completion.

\textbf{M1: Unseen Chain Check.} If $(o_i, c_i) \notin \mathcal{D}$, the trace is flagged as a path anomaly without requiring model inference.

\textbf{M2: Path Prediction Check.} Given the history window, the trace is flagged if $\boldsymbol{\pi}_i[k_i] < \tau_p$, i.e., the softmax probability assigned to the actual chain ID falls below a threshold. We use $\tau_p = 0.2$. Compared to the top-$g$ criterion used in DeepLog, this soft probability threshold provides more fine-grained discrimination between expected and unexpected transitions.

\textbf{M3: Dual-Channel Latency Check.} We combine two complementary channels: (A)~a model-based check that flags events where the actual-to-predicted latency ratio $l_i / \hat{l}_i^{\text{raw}} > \tau_r$, leveraging the LSTM's learned chain-conditional expectations; and (B)~a per-chain percentile fallback that flags $l_i$ exceeding the $q$-th training percentile, providing robustness at trace boundaries and for rare chains. An event is flagged if either channel fires.

\textbf{M4: Trace Length Check.} At trace completion, we flag traces whose span count falls outside the per-root-endpoint $[\min, \max]$ range observed in training, catching span-deletion anomalies in short traces where surviving spans are individually normal.

A trace is reported as anomalous if any mechanism fires. M1--M3 operate online; M4 operates at trace completion as a structural safety net. We view M3 and M4 as lightweight safeguards for common latency and length anomalies; the main representation-level contribution lies in M1 and M2, which directly arise from chain-aware encoding.


\begin{figure}[t]
\centering
\resizebox{\columnwidth}{!}{%
\begin{tikzpicture}[
    >=Stealth,
    node distance=0.35cm,
    box/.style={draw, rounded corners=2pt, minimum height=0.6cm,
        font=\small, align=center, inner sep=3pt},
    mbox/.style={box, fill=red!8, minimum width=1.5cm},
    arr/.style={->, thick, black!70},
    every node/.style={font=\small},
    sub/.style={font=\footnotesize, text=black!50},
]

\node[box, fill=blue!8, minimum width=1.1cm] (e1) {\texttt{preserve}};
\node[box, fill=blue!8, minimum width=1.1cm, right=0.3cm of e1] (e2) {\texttt{prsv/GET}};
\node[box, fill=blue!15, minimum width=1.1cm, right=0.3cm of e2] (e3) {\texttt{sec/check}};

\draw[arr, black!40] (e1.east) -- (e2.west);
\draw[arr, black!40] (e2.east) -- (e3.west);

\node[font=\footnotesize, text=black!45, above=0.06cm of e2] {sliding window (trace spans)};

\node[mbox, right=0.75cm of e3, font=\small\bfseries,
      minimum width=1.4cm, minimum height=0.75cm] (m1)
    {\textbf{M1}\\[-1pt]{\footnotesize\color{black!50} chain $\notin \mathcal{D}$?}};
\node[sub, above=0.01cm of m1, text=red!60, font=\scriptsize] {\emph{no inference}};

\node[box, fill=orange!18, draw=orange!60, line width=0.8pt,
      below=0.55cm of e2, minimum width=3.8cm,
      minimum height=1.2cm, inner sep=4pt] (enc)
    {\textbf{Chain-Aware Encoding}\\[2pt]
     {\footnotesize\color{black!50} \texttt{(sec/check, [prsv$\to$prsv/GET$\to$sec/check])} $\to$ $k_i \in \mathcal{D}$}\\[1pt]
     {\footnotesize\color{black!50} $\mathbf{x} = [\mathbf{E}(k) \;\|\; \tilde{l} \;\|\; \tilde{\Delta} \;\|\; p]$}};

\node[box, fill=green!10, below=0.35cm of enc, minimum width=3.8cm,
      font=\small\bfseries] (lstm) {2-Layer LSTM};

\node[box, fill=yellow!12, below left=0.4cm and 0.0cm of lstm,
      minimum width=1.5cm, minimum height=0.75cm] (ph)
    {Path Head\\[-1pt]{\footnotesize\color{black!50} next-event prob.}};
\node[box, fill=yellow!12, below right=0.4cm and 0.0cm of lstm,
      minimum width=1.5cm, minimum height=0.75cm] (lh)
    {Latency Head\\[-1pt]{\footnotesize\color{black!50} predicted lat.}};

\node[mbox, below=0.3cm of ph, font=\small\bfseries,
      minimum height=0.75cm] (m2)
    {\textbf{M2} $p < \tau_p$?\\[-1pt]{\footnotesize\color{black!50} path anomaly}};
\node[mbox, below=0.3cm of lh, font=\small\bfseries,
      minimum height=0.75cm] (m3)
    {\textbf{M3} dual-ch.\\[-1pt]{\footnotesize\color{black!50} latency anom.}};

\node[sub, text=blue!50, font=\scriptsize, left=0.1cm of m2, rotate=90, anchor=south] {\emph{online (M1--M3)}};

\coordinate (mid) at ($(m2.south)!0.5!(m3.south)$);
\node[mbox, below=0.4cm of mid, font=\small\bfseries,
      minimum height=0.75cm] (m4)
    {\textbf{M4} length\\[-1pt]{\footnotesize\color{black!50} post-trace}};
\node[box, fill=gray!15, below=0.35cm of m4,
      font=\small\bfseries, minimum width=2.2cm] (out) {Normal / Anomaly};


\draw[arr] (e1.south) -- (e1.south |- enc.north);
\draw[arr] (e2.south) -- (enc.north);
\draw[arr] (e3.south) -- (e3.south |- enc.north);

\draw[arr, dashed, red!50] (e3.east) -- (m1.west)
    node[midway, above, font=\scriptsize] {$\mathcal{D}$};

\draw[arr] (enc) -- (lstm);

\draw[arr] ([xshift=-0.45cm]lstm.south) -- (ph.north);
\draw[arr] ([xshift=0.45cm]lstm.south)  -- (lh.north);

\draw[arr] (ph) -- (m2);
\draw[arr] (lh) -- (m3);

\draw[arr] (m2.south) -- ++(0,-0.1) -| ([xshift=-0.12cm]m4.north);
\draw[arr] (m3.south) -- ++(0,-0.1) -| ([xshift=0.12cm]m4.north);

\draw[arr] (m4) -- (out);

\coordinate (m1drop) at ([yshift=-0.3cm]m1.south);
\coordinate (m1right) at ([xshift=0.4cm]m1drop -| m1.east);
\draw[arr, dashed, red!50]
    (m1.south) -- (m1.south |- out.east) -- (out.east);

\end{tikzpicture}%
}
\caption{\method architecture. The chain-aware encoding (highlighted) replaces endpoint-only vocabulary with (endpoint, invocation chain) pairs, enabling out-of-dictionary detection (M1, dashed) without model inference. M2--M3 operate per-event using LSTM predictions; M4 checks trace structure at completion.}
\label{fig:architecture}
\end{figure}
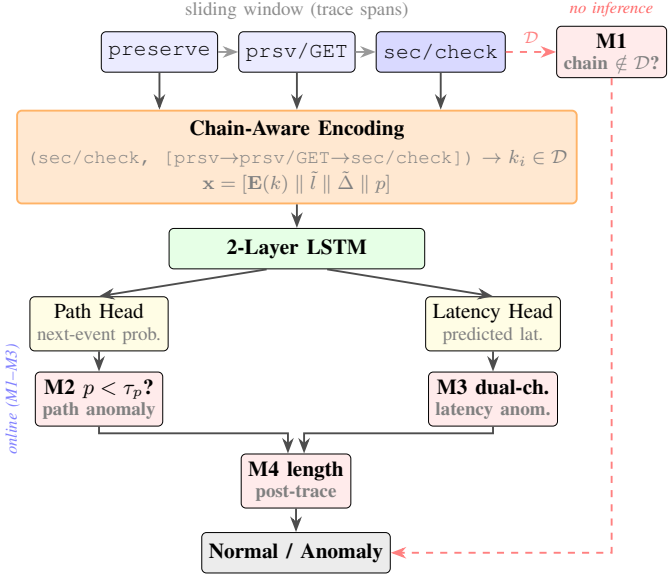


\section{Experiments}
\label{sec:exp}

We evaluate \method on the TrainTicket benchmark to answer three research questions:

\begin{itemize}
    \item \textbf{RQ1}: How does \method compare with representative baselines under the controlled evaluation setting?
    \item \textbf{RQ2}: What is the contribution of chain-aware encoding, and how sensitive is \method to the window size?
    \item \textbf{RQ3}: What types of anomalies does \method detect that endpoint-only approaches miss, and what is the underlying mechanism?
\end{itemize}

\subsection{Dataset and Setup}
\label{subsec:setup}

\textbf{Dataset.} We use the TrainTicket~\cite{trainticket} benchmark (41 services) with the publicly available trace dataset from Zenodo~\cite{eadro}. After reconstructing traces by \texttt{traceID}, we obtain 9{,}200 normal traces (54{,}326 spans, median 3 spans per trace, 23 distinct endpoint labels, $|\mathcal{D}|{=}59$ chain-aware tokens). For brevity, \texttt{ts-X-service} is shortened to \texttt{X} in case studies.

\textbf{Train/Validation/Test split.} Normal traces are split 70/15/15 into training (6{,}440), validation (1{,}380), and test (1{,}380) sets. The validation set serves two purposes: (1)~early stopping during model training (normal traces only), and (2)~threshold tuning for detection hyperparameters, where we inject anomalies into copies of validation normals (with a separate random seed) to form a labeled validation set. The test set is used only once for final evaluation.

\textbf{Anomaly injection.} Following standard practice~\cite{traceanomaly}, we inject two categories of anomalies into copies of the held-out test normals, each producing a paired variant of the source trace:

\emph{Latency anomalies (690 traces):} For a randomly selected span, we multiply its \texttt{duration} by a factor uniformly sampled from $[10, 50]$ and propagate the added latency upward to all ancestor spans, simulating service slowdowns with realistic cascading effects.

\emph{Path anomalies (690 traces):} We apply one of three structural mutations:
(i)~\emph{span deletion}---remove a non-root span and reparent its children to preserve a valid trace tree (only for traces with $>$4 spans to ensure detectability);
(ii)~\emph{span swap}---exchange the timestamps of two distant, non-sibling spans from different subtrees;
(iii)~\emph{span duplication}---insert a copy of a span at a non-adjacent position.
The test set thus contains 1{,}380 normal + 1{,}380 anomaly = 2{,}760 traces.

\subsection{Baselines}
\label{subsec:baselines}

We compare against four baselines: \textbf{Rule-based} ($\mu + 3\sigma$ latency threshold per endpoint), \textbf{Isolation Forest} (7-dimensional trace-level features, \texttt{contamination}$=$0.05), \textbf{Autoencoder} (same features, reconstruction error threshold at the 95th percentile), and \textbf{DeepLog}~\cite{deeplog} (same LSTM architecture and hyperparameters as \method, but with endpoint-only vocabulary, top-$g$ path detection, and per-endpoint latency percentile---no dual-channel latency, no position encoding, no trace length check). DeepLog is the primary comparison for evaluating chain-aware encoding. All hyperparameters ($g$, percentile level) are selected on the validation set by grid search.

\subsection{Implementation Details}
\label{subsec:impl}

We use $h{=}2$, $d_e{=}32$, $d_h{=}64$, a 2-layer LSTM (${\sim}$65K parameters), trained with Adam (lr$=$1e-3, batch$=$128, early stopping patience$=$5, $\alpha{=}1.0$). Detection thresholds are selected on the validation set: for \method, the latency percentile $q{=}96.5$ is chosen by grid search over $\{95, 95.5, \ldots, 99\}$ to maximize F1, with $\tau_p{=}0.2$ and $\tau_r{=}5$ fixed. For DeepLog, $g{=}2$ and $q{=}97.5$ are jointly selected from $g \in \{1,2,3,5,10\}$ and the same percentile grid. Per-event inference takes ${\sim}$0.6\,ms on a single CPU core (averaged over 1{,}000 runs, batch size 1).

\subsection{Main Results (RQ1)}
\label{subsec:main}

Table~\ref{tab:main} reports detection performance. \method achieves the highest F1 among the evaluated baselines (94.3\%), outperforming our DeepLog reimplementation by 5.3 percentage points. The improvement is concentrated in recall (+12.0pp), reflecting \method's ability to detect path anomalies that endpoint-only prediction misses (path recall: 99.1\% vs.\ 75.1\%). Latency recall is comparable between the two methods (97.5\% vs.\ 97.7\%), confirming that the gap is driven by representation granularity rather than the latency detection mechanism. Traditional baselines (Rule-based, Isolation~Forest, Autoencoder) operate on aggregate features and cannot exploit sequential call structure, leading to substantially lower performance.

\begin{table}[t]
\centering
\caption{Anomaly detection results on TrainTicket. Detection thresholds for \method and DeepLog are selected on the validation set. Best results in bold.}
\label{tab:main}
\small
\begin{tabular}{@{}lcccccc@{}}
\toprule
Method & Prec. & Rec. & F1 & Lat-R & Path-R \\
\midrule
Rule-based              & 78.3 & 7.3  & 13.4 & 12.0 & 2.6 \\
Isolation Forest        & 92.8 & 53.3 & 67.7 & 99.0 & 7.7 \\
Autoencoder             & 83.3 & 0.7  & 1.4  & 1.2  & 0.3 \\
DeepLog~\cite{deeplog}  & 91.8 & 86.4 & 89.0 & 97.7 & 75.1 \\
\textbf{\method (ours)} & \textbf{90.7} & \textbf{98.3} & \textbf{94.3} & 97.5 & \textbf{99.1} \\
\bottomrule
\end{tabular}
\end{table}

\subsection{Ablation and Sensitivity (RQ2)}
\label{subsec:ablation}

\textbf{Encoding ablation.} To isolate chain-aware encoding, we train an endpoint-only model with the \emph{same} architecture and \emph{same} M1--M4 detection pipeline, differing only in vocabulary (23 endpoint labels vs.\ 59 chain-aware tokens). This controlled ablation achieves 93.2\% F1 (path recall 96.2\%), compared with 94.3\% F1 (path recall 99.1\%) for \method. Although the aggregate F1 gap is narrowed by overlapping OR-based detection mechanisms and TrainTicket's shallow traces (median 3 spans), chain-aware encoding reduces missed path anomalies from 26 to 6 (of 690) and false positives from 155 to 140. Removing M4 or the latency loss ($\alpha{=}0$) changes F1 by ${\leq}0.07$pp, indicating that the detector is robust to these auxiliary components. Section~\ref{subsec:case} analyzes the prediction distributions behind the remaining gap.

\textbf{Window size sensitivity.} F1 is robust across $h \in \{1, \ldots, 5\}$ (within $0.16$pp); we use $h{=}2$ because the median trace has only 3 spans.

\subsection{Where Does Chain Awareness Help? (RQ3)}
\label{subsec:case}

Because aggregate F1 mixes latency and path anomalies, we analyze the 166 anomaly traces detected by \method but missed by DeepLog. Among these, 161 path-related cases fall into two categories, each illustrating one consequence of chain-aware encoding; the remaining 5 are latency-boundary cases resolved by dual-channel detection.

\textbf{Category~A: Out-of-dictionary chains (24 traces).}
The endpoint \texttt{security/check} is normally invoked exclusively through the chain \texttt{preserve}$\,\to\,$\texttt{preserve/GET}$\,\to\,$\texttt{security/check}. When a span deletion removes the intermediate \texttt{preserve/GET}, the resulting invocation chain becomes a novel pair not seen in the training dictionary $\mathcal{D}$. \method flags this immediately via M1, before any model inference. DeepLog, whose vocabulary contains only the endpoint \texttt{security/check}, observes a familiar normal token and has no mechanism to detect the topology change. While unseen chains require care under service evolution (Section~\ref{sec:discuss}), these cases illustrate how the encoding exposes topology changes directly; Category~B examines the in-vocabulary cases where the distributional effect is most visible.

\textbf{Category~B: Sharper context-conditional distributions (137 traces).}
The remaining cases involve anomalies where the endpoint is in-vocabulary but the invocation chain is unusual in the current context. Consider a trace where the sliding window ends with \texttt{travel/queryInfo}$\,\to\,$\texttt{travel/GET}; the actual next event is \texttt{basic/GET}, reached through a 6-hop chain via \texttt{ticketinfo}. DeepLog, predicting over endpoints, assigns \texttt{basic/GET} a probability of $0.99$---the endpoint is highly expected regardless of how it was reached. \method, predicting over (endpoint, chain) pairs, distinguishes \emph{which} chain variant of \texttt{basic/GET} should appear: the actual (anomalous) chain receives $p{=}1.9{\times}10^{-5}$, while a different (normal) chain variant of the same endpoint receives $p{=}0.99$. The probability gap exceeds $10^4$.

Across all 137 such cases, DeepLog assigns the actual endpoint a median probability of $0.88$ (well above any reasonable threshold), while \method assigns the actual (endpoint, chain) pair a median probability of $7 \times 10^{-5}$---a gap of four orders of magnitude (median $12{,}444\times$). In every one of these 137 cases, the endpoint ranks in DeepLog's top-2 predictions. This illustrates a fundamental limitation of top-$g$ endpoint prediction: once the endpoint itself is considered likely, the model has no mechanism to distinguish the anomalous invocation chain from a normal one.

\textbf{Distributional advantage at the representation level.}
Categories~A and~B explain where \method detects anomalies that DeepLog misses. To test whether this advantage is systematic rather than case-specific, we compare \method against the endpoint-only ablation from Section~\ref{subsec:ablation} (identical architecture and pipeline), isolating the effect of vocabulary granularity on prediction distributions.

Across all 571 path-anomaly events where \method's M2 fires, 93 receive a probability above $0.2$ from the endpoint-only ablation---i.e., the endpoint-only model confidently judges these anomalies as normal (median $p{=}0.91$). Under chain-aware encoding, the same events receive median $p{=}0.002$ ($508\times$ lower)---the model's judgment is reversed, not merely sharpened. These are systematic blind spots for endpoint-only prediction, not borderline cases that can be resolved by minor threshold tuning.

\begin{figure}[t]
\centering
\includegraphics[width=\columnwidth]{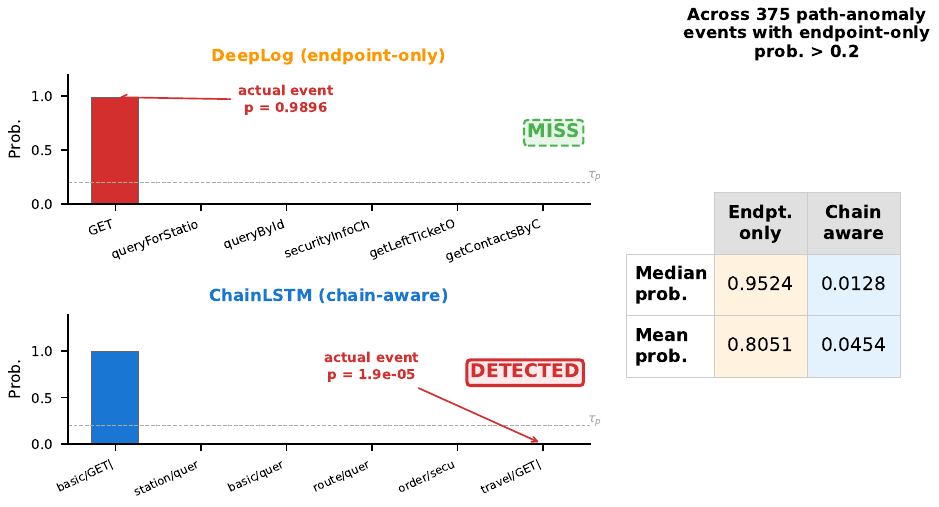}
\caption{Distributional comparison between endpoint-only and chain-aware prediction.
\textbf{Left:} Bars show next-token probabilities for the same event under endpoint-only (top) and chain-aware (bottom) vocabularies; the hatched bar marks the observed event. The comparison shows how a likely endpoint can become an unlikely chain variant.
\textbf{Right:} Each dot is one path-anomaly event scored under both encodings; dotted lines mark $\tau_p{=}0.2$, and the shaded region contains events missed by endpoint-only prediction but detected by \method.}
\label{fig:dist}
\end{figure}

Figure~\ref{fig:dist} (left) illustrates a representative case. A span-swap mutation reorders two spans within a trace; the anomalous span's endpoint is \texttt{GET}, which is the most common next-event prediction in this context. The endpoint-only model assigns $p{=}0.998$ to \texttt{GET}---indistinguishable from a normal event. \method, however, splits \texttt{GET} into multiple chain-aware tokens depending on the invocation path. The actual (anomalous) chain token receives $p{=}1.1{\times}10^{-5}$, while a different chain variant of the same endpoint receives most of the probability mass. The anomaly is invisible at endpoint granularity but becomes a clear outlier at chain granularity.

Figure~\ref{fig:dist} (right) shows that this is not an isolated example: across all 93 blind-spot events, the dots are concentrated far below the diagonal, indicating that chain-aware encoding systematically transforms high-confidence endpoint predictions into low-probability chain predictions. This provides a wider safety margin across a broad range of thresholds.

\section{Discussion}
\label{sec:discuss}

\method demonstrates that chain-aware encoding exposes structural signals not directly represented by endpoint-only encodings, while retaining real-time per-event operation. Beyond the 166-trace detection gap, the distributional analysis (Figure~\ref{fig:dist}) shows that chain-aware encoding systematically transforms a large class of path anomalies from distributional blind spots into clear outliers, shifting median prediction probability from $0.91$ to $0.002$.

\textbf{When does chain-aware encoding help?} The encoding's advantage depends on invocation diversity: it is most effective when the same endpoint is reached through multiple distinct chains with different normal behaviors. In flat API gateway architectures where all calls route through a single proxy, chains collapse to depth one and the encoding degenerates to endpoint-only. TrainTicket, with $|\mathcal{D}|{=}59$ tokens over 23 endpoint labels, exhibits moderate diversity; production systems with deeper call graphs may show larger gains.

\textbf{Dictionary evolution.} A key practical question is distinguishing \emph{new normal chains} (from service evolution) from \emph{anomalous chains}. DeepLog treats all unseen tokens identically. Chain-aware encoding enables a richer strategy: when a new (endpoint, chain) pair appears, detection can evolve from a hard unseen-chain rejection to a soft decision by marginalizing over all dictionary entries that share the same endpoint, $\sum_{k: \text{ep}(k)=o_{\text{new}}} \boldsymbol{\pi}_i[k]$. A high marginalized probability suggests the endpoint is expected in context and the new chain reflects legitimate evolution; a low score signals a true anomaly. This two-level judgment is unavailable to flat-vocabulary approaches. The dictionary is constructed in a single $O(n)$ pass; for production systems where $|\mathcal{D}|$ may reach $10^3$--$10^4$, this remains within LSTM embedding capacity.

\textbf{Encoding granularity trade-off.} Full invocation chains represent one end of a design space. Caller-callee pairs (immediate parent only) yield smaller dictionaries but lose deep context; n-gram path models offer an intermediate granularity. We chose full chains because microservice traces are typically shallow (median 3 spans in TrainTicket), keeping $|\mathcal{D}|$ manageable. In systems with deep or highly variable call graphs, truncated chains or hierarchical encodings may provide a better trade-off between context richness and dictionary size.

\textbf{Limitations and future work.} (i)~Our evaluation uses one benchmark, synthetic anomalies, and a balanced test set; thus, its precision and F1 may not reflect false-alarm burden at production anomaly rates. (ii)~The injected path mutations are designed to exercise structural sensitivity and may not capture previously unseen but legitimate chains introduced by normal service evolution. The dictionary-evolution strategy above outlines how endpoint-marginalized scoring can soften unseen-chain rejection, but this strategy remains to be validated under continuous production evolution. (iii)~We have not evaluated naturally occurring incidents or compared against intermediate context representations such as caller-callee pairs. Future work includes validation on production traces, adaptive dictionary updates for legitimate new chains, exploring the design space between endpoint-only and full-chain encoding (e.g., caller-callee pairs, n-gram paths, and hierarchical encodings), and extending chain-aware encoding to other sequential invocation settings. Related sequential decision research likewise highlights the value of representing context explicitly. Adaptive pressure incorporates upstream conditions in traffic control~\cite{duan_bayesian_traffic}, while dynamic context selection and environment-aware interaction modeling support multi-agent learning and real-time trajectory prediction~\cite{duan_adaptive_context,duan_mavent}.

\section{Conclusion}
\label{sec:conclusion}

We proposed \method, a lightweight online detector that encodes trace events as (endpoint, invocation chain) pairs. On TrainTicket, it achieves 94.3\% F1, outperforming DeepLog by 5.3 points with comparable latency recall and 99.1\% path recall. Case analysis shows that invocation context exposes structural anomalies hidden by endpoint-only prediction, motivating broader evaluation of chain-aware encoding.

\section*{Acknowledgment}
OpenAI ChatGPT and Codex assisted with language editing, TikZ figures, and auxiliary experimental scripts (e.g., result formatting, plotting); all outputs were reviewed by the authors.

\bibliographystyle{IEEEtran}
\bibliography{references}

@inproceedings{deeplog,
  author    = {Du, Min and Li, Feifei and Zheng, Guineng and Srikumar, Vivek},
  title     = {{DeepLog}: Anomaly Detection and Diagnosis from System Logs through Deep Learning},
  booktitle = {Proc. ACM Conference on Computer and Communications Security (CCS)},
  pages     = {1285--1298},
  year      = {2017}
}

@inproceedings{traceanomaly,
  author    = {Liu, Ping and Xu, Haowen and Ouyang, Qianyu and Jiao, Rui and Chen, Zhekang and Zhang, Shenglin and Yang, Jiahai and Mo, Linlin and Zeng, Jice and Xue, Wenman and Pei, Dan},
  title     = {Unsupervised Detection of Microservice Trace Anomalies through Service-Level Deep {Bayesian} Networks},
  booktitle = {Proc. IEEE International Symposium on Software Reliability Engineering (ISSRE)},
  pages     = {48--58},
  year      = {2020}
}

@inproceedings{ticad,
  author    = {Du, Qingfeng and Zhao, Liming and Tian, Fei and Han, Yanghui},
  title     = {Trace-Based Anomaly Detection with Contextual Sequential Invocations},
  booktitle = {Proc. International Conference on Database and Expert Systems Applications (DEXA)},
  series    = {LNCS},
  pages     = {95--109},
  year      = {2023}
}

@article{lstm,
  author  = {Hochreiter, Sepp and Schmidhuber, J{\"u}rgen},
  title   = {Long Short-Term Memory},
  journal = {Neural Computation},
  volume  = {9},
  number  = {8},
  pages   = {1735--1780},
  year    = {1997}
}

@inproceedings{loganomaly,
  author    = {Meng, Weibin and Liu, Ying and Zhu, Yichen and Zhang, Shenglin and Pei, Dan and Liu, Yuqing and Chen, Yihao and Zhang, Ruizhi and Tao, Shimin and Sun, Pei and Zhou, Rong},
  title     = {{LogAnomaly}: Unsupervised Detection of Sequential and Quantitative Anomalies in Unstructured Logs},
  booktitle = {Proc. International Joint Conference on Artificial Intelligence (IJCAI)},
  pages     = {4739--4745},
  year      = {2019}
}

@inproceedings{logrobust,
  author    = {Zhang, Xu and Xu, Yong and Lin, Qingwei and Qiao, Bo and Zhang, Hongyu and Dang, Yingnong and Xie, Chunyu and Yang, Xinsheng and Cheng, Qian and Li, Ze and others},
  title     = {Robust Log-Based Anomaly Detection on Unstable Log Data},
  booktitle = {Proc. ACM Joint Meeting on European Software Engineering Conference and Symposium on the Foundations of Software Engineering (ESEC/FSE)},
  pages     = {807--817},
  year      = {2019}
}

@article{tracegra,
  author  = {Chen, Jian and Liu, Fagui and Jiang, Jun and Zhong, Guoxiang and Xu, Dishi and Tan, Zhuanglun and Shi, Shangsong},
  title   = {{TraceGra}: A Trace-based Anomaly Detection for Microservice Using Graph Deep Learning},
  journal = {Computer Communications},
  volume  = {204},
  pages   = {109--117},
  year    = {2023}
}

@article{dapper,
  author  = {Sigelman, Benjamin H. and Barroso, Luiz Andr{\'e} and Burrows, Mike and Stephenson, Pat and Plakal, Manoj and Beaver, Donald and Jaspan, Saul and Shanbhag, Chandan},
  title   = {Dapper, a Large-Scale Distributed Systems Tracing Infrastructure},
  journal = {Google Technical Report},
  year    = {2010}
}

@inproceedings{opentelemetry,
  author    = {Boten, Austin Parker and Morgan McLean and Daniel Dyla},
  title     = {{OpenTelemetry}: An Open Standard for Distributed Tracing},
  booktitle = {Proc. KubeCon + CloudNativeCon},
  year      = {2019}
}

@inproceedings{transformer,
  author    = {Vaswani, Ashish and Shazeer, Noam and Parmar, Niki and Uszkoreit, Jakob and Jones, Llion and Gomez, Aidan N. and Kaiser, {\L}ukasz and Polosukhin, Illia},
  title     = {Attention Is All You Need},
  booktitle = {Proc. Advances in Neural Information Processing Systems (NeurIPS)},
  pages     = {5998--6008},
  year      = {2017}
}

@article{aiops_survey,
  author  = {Notaro, Paolo and Cardoso, Jorge and Gerndt, Michael},
  title   = {A Systematic Mapping Study in {AIOps}},
  journal = {Proc. International Conference on Service-Oriented Computing (ICSOC)},
  pages   = {110--123},
  year    = {2020}
}

@article{trainticket,
  author  = {Zhou, Xiang and Peng, Xin and Xie, Tao and Sun, Jun and Ji, Chao and Li, Wenhai and Ding, Dan},
  title   = {Fault Analysis and Debugging of Microservice Systems: Industrial Survey, Benchmark System, and Empirical Study},
  journal = {IEEE Transactions on Software Engineering},
  volume  = {47},
  number  = {2},
  pages   = {243--260},
  year    = {2021}
}

@inproceedings{eadro,
  author    = {Lee, Cheryl and Yang, Tianyi and Chen, Zhuangbin and Su, Yuxin and Yang, Yongqiang and Lyu, Michael R.},
  title     = {{Eadro}: An End-to-End Troubleshooting Framework for Microservices on Multi-source Data},
  booktitle = {Proc. IEEE/ACM International Conference on Software Engineering (ICSE)},
  pages     = {1750--1762},
  year      = {2023}
}

@article{duan_bayesian_traffic,
  author  = {Duan, Wenchang and Gao, Zhenguo and He, Jiwan and Xian, Jinguo},
  title   = {Bayesian Critique-Tune-Based Reinforcement Learning With Adaptive Pressure for Multi-Intersection Traffic Signal Control},
  journal = {IEEE Transactions on Intelligent Transportation Systems},
  volume  = {26},
  number  = {10},
  pages   = {14968--14983},
  year    = {2025},
  doi     = {10.1109/TITS.2025.3581858}
}

@inproceedings{duan_adaptive_context,
  author    = {Duan, Wenchang and Yu, Yaoliang and He, Jiwan and Shi, Yi},
  title     = {Adaptive Context Length Optimization with Low-Frequency Truncation for Multi-Agent Reinforcement Learning},
  booktitle = {Advances in Neural Information Processing Systems},
  volume    = {38},
  year      = {2025}
}

@article{duan_mavent,
  author  = {Duan, Wenchang},
  title   = {{MAVEN-T}: Multi-Agent enVironment-Aware Enhanced Neural Trajectory Predictor with Reinforcement Learning},
  journal = {arXiv preprint arXiv:2604.10169},
  year    = {2026}
}

\end{document}